\documentclass[11pt,twoside]{article}
\usepackage{asp2014}

\aspSuppressVolSlug
\resetcounters

\begin{document}

\title{CHIPP: INAF pilot project for HTC, HPC and HPDA}

\author{Giuliano~Taffoni,$^1$ Ugo~Becciani,$^2$ Bianca~Garilli,$^3$ Gianmarco~Maggio,$^1$ Fabio~Pasian,$^1$ Grazia~Umana,$^2$ Riccardo~Smareglia,$^1$ and Fabio~Vitello$^2$}
\affil{$^1$INAF - OATs, Trieste, Friuli Venezia Giulia, Italy; \email{giuliano.taffoni@inaf.it}}
\affil{$^2$INAF - OACt, Catania, Catania, Italy}
\affil{$^3$INAF - IASF, Milano, Milano, Italy}

\paperauthor{Giuliano~Taffoni}{}{https://orcid.org/0000-0002-4211-6816}{INAF - OATs}{Osservatorio Astronomico di Trieste}{Trieste}{}{34143}{Italy}
\paperauthor{Ugo~Becciani}{ugo.becciani@inaf.it}{https://orcid.org/0000-0002-4389-8688}{INAF- OACt}{Osservatorio Astrofisico di Catania}{Catania}{}{95123}{Italy}
\paperauthor{Bianca~Garilli}{bianca.garilli@inaf.it}{https://orcid.org/0000-0001-7455-8750}{INAF - IASF}{ Istituto di Astrofisica Spaziale e Fisica Cosmica}{Milano}{}{20133}{Italy}
\paperauthor{Gianmarco~Maggio}{gianmarco.maggio@inaf.it}{https://orcid.org/0000-0003-4020-4836}{INAF - OATs}{Osservatorio Astronomico di Trieste}{Trieste}{}{34143}{Italy}
\paperauthor{Fabio~Pasian}{fabio.pasian@inaf.it}{https://orcid.org/0000-0002-4869-3227}{INAF - OATs}{Osservatorio Astronomico di Trieste}{Trieste}{}{34143}{Italy}
\paperauthor{Grazia~Umana}{grazia.umana@inaf.it}{https://orcid.org/0000-0002-6972-8388}{INAF- OACt}{Osservatorio Astrofisico di Catania}{Catania}{}{95123}{Italy}
\paperauthor{Riccardo~Smareglia}{riccardo.smareglia@inaf.it}{https://orcid.org/0000-0001-9363-3007}{INAF - OATs}{Osservatorio Astronomico di Trieste}{Trieste}{}{34143}{Italy}
\paperauthor{Fabio~Vitello}{fabio.vitello@inaf.it}{https://orcid.org/0000-0003-2203-3797}{INAF- OACt}{Osservatorio Astrofisico di Catania}{Catania}{}{95123}{Italy}




  
\begin{abstract}
CHIPP 
(Computing HTC in INAF Pilot Project)
is an Italian project funded by the Italian Institute for Astrophysics (INAF) and promoted by the ICT office of INAF. 
The main purpose of the CHIPP project is to coordinate the use of, and access to, already existing high throughput computing and high--performance 
computing and data processing resources (for small/medium size programs) for the INAF community.  Today, Tier--2/Tier--3 systems (1,200 CPU/core) are 
provided at the INAF institutes at Trieste and Catania, but in the future, the project will evolve including also other computing infrastructures.
During the last two years more than 30 programs have been approved for a total request of 30 Million CPU-h. Most of the programs are HPC, data reduction and analysis, machine learning.
In this poster, we describe in details the CHIPP infrastructures and the results of the first two years of activity.
\end{abstract}

\section{Introduction}
Astronomy and Astrophysics (AA) is one of the research areas in Physics that needs more and more high performing software and infrastructures, as the necessity for supercomputers arises both in the observational and theoretical realms of AA. 
The amount of data produced by the upcoming generation of surveys and scientific instruments (e.g.  Square Kilometer Array,  Cherenkov Telescope Array,  Extremely Large Telescope E-ELT,  James Webb Space Telescope,  Euclid satellite, and eROSITA All-Sky Survey), needs more and more resources to be efficiently post-processed, analyzed and stored.
Moreover, numerical simulations, a theoretical counterpart capable of reproducing the formation and evolution of the cosmic structures of our Universe, must reach both larger volumes and higher resolutions to cope with a large amount of data of the upcoming surveys. 

In the last decade, the National Institute of Astrophysics in Italy (INAF)  contributed to a number of HPC and HTC projects  (e.g. ExaNeSt \citep{KATEVENIS201858, 8049832}, EuroExa, EGI-Engage \citep{BERTOCCO201836}, EOSC-Pilot, DHTCS-IT).  
Moreover, INAF recognized the importance to offer computing facilities of medium size to its user community and to invest in developing know-how in computing and data analytic. 

In this paper, we present the experiment to coordinate the use of, and access to,  high throughput computing (HTC) and high-performance computing (HPC) and data processing resources.

\section{Scope and objective of the project}
CHIPP is a coordinated action supported by INAF and based on existing infrastructure and benefit of the know-how offered by the involved centres.   CHIPP main objectives are:
\begin{itemize}
    \item  Investigate to what extent the Italian astronomical community is interested in using a distributed computing infrastructure;
    \item Investigate whether the infrastructure meets the Astronomers requirements and collect comment and suggestions to optimize it;
    \item Investigate the best approach to offer the resources (e.g. periodic call, rolling based, on-demand access)
    \item Implement efficient user support and investigate which support level is required (system management, support for software development, support for optimization and debugging)
\end{itemize}

This project is the prototype and test case for the development of an extensive distributed computing infrastructure for astronomers able to offer multiple resources: HPC, HTC, Cloud.

\section{CHIPP computing resources}
The CHIPP project benefits of two computing centres located at  INAF--OATs and INAF--OACt.

The HOTCAT Cluster at INAF--OATs \citep{P6.7_adassxxix} is an HPC infrastructure with 1400 INTEL Haswell E5-4627v3 cores with 6GB RAM per Core (8.5TB RAM total) and 500 TB parallel storage based on BeeGFS. The interconnect is Infiniband ConnectX -3 Pro Dual QSFP+ 54Gbs. Beside the HPC cluster, INAF--OATs is also providing an OpenStack based cloud stack with 200INTEL Westmere E5620 cores with 8GB RAM per Core and 75 TB Swift ObjectStorage.

INAF--OACt MUP cluster is a HTC infrastructure with 384 INTEL E5-2620V2  with  5.2GB RAM per Core (1TB RAM total) and  70 TB parallel storage (NFS).  The interconnect is a 10 Gbs Ethernet network.

All the computing infrastructures are equipped with more than 60 software environment for Astronomical data reduction and analysis. They offer tools for software development, profiling and debugging.

\section{CHIPP Implementation}
CHIPP pilot is coordinating the services offered by the two computing centres, and it is verifying that the INAF community is profiting successfully of the computing resources. 

The computing centres provide the system support and the installation and maintenance of libraries and tools. During this first implementation of the pilot project, the computing centres are not providing any support to optimize and debug users codes. 
Due to a large number of requests for this type of service,  in the new implementation of the project, we will extend our support towards software optimization and debugging.

CHIPP collects the requirements and suggestions from the users to improve the platforms.  Computing centres provide customized environments based on the collected requirements. 

\begin{table}[!ht]
\caption{The results of the four calls for proposals. For each call we present the number of submitted proposals, the number of accepted proposals, the total amount of core hours reserved for all the projects. We report also the completion rate, i.e the percentage of core hours actually consumed.}
\smallskip
\begin{center}
{\small
\begin{tabular}{lcccc}
\tableline
\noalign{\smallskip}
            & I Call     & II Call & III Call & IV call  \\
\noalign{\smallskip}   
\tableline
\noalign{\smallskip}
Date &  15 May 2017 & 15 Jan 2018 & 30 July 2018 & 30 May 2019 \\
\# of proposals         & 25         & 14      & 9        & 13       \\
\# of proposal accepted & 16         & 14      & 9        & 13       \\
Core hours allocated    & $2.5x10^6$ & $1.8x10^6$ & $1.2x10^6$  & $1.4x10^6$  \\
Completion rate         & 33\%       & 50\%    & 55\%     & on-going \\
\noalign{\smallskip}
\tableline\
\end{tabular}
}
\end{center}
\label{table1}
\end{table}

The main CHIPP activity is promoting periodic competitive calls for computing resources, one every 6 months for large projects (~80000 core hours) and rolling open calls for small projects (<10000 core hours) of one month. Projects supported by CHIPP are chosen based on innovation potential, scientific excellence and relevance criteria.

In Table~\ref{table1}, we show the results of the 4 call for proposals. 
The number of proposals presented during the first call was extremely high compared to the others; then it is steadily around 10 to 15  proposals each call.
According to our experience,  that corresponds to greater user awareness of what a computing resource can offer in terms of capabilities and environment.

During the project, we also modify the way a user can ask for computing resources, 
thanks to the introduction of the rolling open calls. 
This kind of calls allows astronomers with an un-planned need of computing resources to access to the infrastructure efficiently.

\section{Conclusion and future activities}

The INAF CHIPP pilot project is offering a distributed computing and storage resources to astronomers and projects on a call based approach. 

CHIPP is a successful experiment that aims to help grow a community of experts in numerical AA and consolidate know--how in usage, maintenance and 
support of computing resources.

The use of distributed computing infrastructure instead of a single computing centre is an architectural choice, in line with the architecture of the  European Open Science Cloud. That allows us to experiment with the operations and services activities necessary in more significant projects as the SKA Regional Center networks.

The small or middle-sized resources offered in CHIPP, are excellent for developing and debugging codes, medium-size production activities, data analysis, and according to our experience, they respond to a specific community requirement,  not covered by supercomputing centres. 
CHIPP resources can also be beneficial for small (and large) projects during their start-up when they need computing resources that are not offered by the project yet.

CHIPP is a successful experiment to define requirements, investigate the type of use and necessities of a broad astronomical community. 
It is offering support to astronomers for the use of the resources and building science platforms based singularity containers, Jupyter notebooks and/or remote desktop. 

The experience gained during the project and the requisites collected are the foundations for the design and development of an extensive distributed computing infrastructure for astronomy in Italy (Tier-1 like): the Data-Star project.

\bibliography{P6-8}


\end{document}